\documentclass{ws-ijmpa}

\begin{document}

\markboth{W. Sch\"afer}
{High $\bf{p}$--forward pions}

\catchline{}{}{}{}{}
%
%%%%%%%%%%%%%%%%%%%%%%%%%%%%%%%%%%%%%%%%%%%%%%%%%%%%%%%%%%%%%%%%%%%%

\title{ FORWARD PION  PRODUCTION AT LARGE TRANSVERSE MOMENTA
   IN PP COLLISIONS AND BEYOND
}

\author{W. SCH\"AFER
%\footnote{
%Typeset names in 8 pt roman, uppercase. Use the footnote to indicate the
%present or permanent address of the author.}
}

\address{Institute of Nuclear Physics,\\ 
ul. Radzikowskiego 152,\\
PL-31-342 Cracow, Poland \\
%\footnote{State completely without abbreviations, the
%affiliation and mailing address, including country. Typeset in 8 pt
%italic.}\\
wo.schaefer@fz-juelich.de}

\maketitle

\begin{history}
\received{Day Month Year}
\revised{Day Month Year}
\end{history}

\begin{abstract}
The inclusive production of high-$p_\perp$ particles (pions) in the beam fragmentation 
regions of high--energy hadronic collisions is driven by the breakup of
valence constituents of the beam hadrons into their two--body
Fock--state components,and their subsequent fragmentation. 
We briefly discuss an approach, that allows the consistent inclusion 
of intrinsic and radiatively generated transverse momenta of initial
state partons,  and describe an extension of our approach to
nuclear targets.

\end{abstract}

\ccode{PACS numbers: 13.85.-t,13.87.-a}

\section{Linear $k_\perp$--factorization}

Inclusive particle production at large transverse momenta $p_\perp$ is
conventionally treated in the collinear factorization approach, well
known from textbooks \cite{Collinear}. A crucial assumption is that $p_\perp$ 
is the only large dimensionful scale, and the Bjorken-$x$'s of
colliding partons are not too small. The differential cross section
takes a factorized form, schematically 
\begin{equation}
d\sigma(pp \to M X) \sim n_i \otimes n_j \otimes d\hat{\sigma}(ij \to
cd) \otimes D_{c \to M} \, ,
\end{equation}
and is calculable in terms of collinear parton densities $n_{i,j}$, extracted,
e.g. from deep inelastic scattering, the fragmentation
functions as extracted, e.g. from $e^+ e^-$ annihilation, and the parton
level cross section calculated in pQCD.  Here, at leading order 
the parton level final state is a back--to--back dijet system.
At higher energies, when perturbatively
large transverse momenta may satisfy $\Lambda_{QCD} \ll p_\perp \ll
\sqrt{s}$, however the collinear factorization is inadequate, and an 
explicit inclusion of transverse momentum degrees of freedom for
initial state partons is called for. The pertinent formalism -- (linear)
$k_\perp$ factorization, many applications and some open problems are 
described in Refs. (\refcite{Ktfac,Antoni}).
In the problem of interest here, production of particles at large
(pseudo)--rapidities in the beam fragmentation region, the relevant 
target partons will be dominantly gluons, carrying small $x$ (see
e.g. Ref. (\refcite{Antoni})). The dominant beam partons will clearly be
valence quarks, and indeed the parton level production process 
may be seen as a scattering of valence quarks into large $p_\perp$
at a Born level, and as a breakup into their quark--gluon component
at the level of the first radiative correction \cite{SingleJet,RealVirtual}.
The distribution of valence quarks is a steep function of their momentum
fraction $x$, as $x\to 1$, and the production of high $p_\perp$ forward
partons must be sensitive to their energy loss in the $q \to qg$
transition. Evaluation of the $p_\perp$--dependent energy loss requires
a calculation of virtual radiative corrections to the radiationless Born
term within the framework of $k_\perp$ factorization, which has been
accomplished in Ref. (\refcite{RealVirtual}). We summarize some of its
salient features. 
First, the parton level cross
section is a linear probe of the target unintegrated gluon density,
emission of slow gluons is consistently absorbed into the BFKL evolution 
of the target unintegrated glue, second the produced $qg$ dijet system
in the breakup is {\emph not} back--to--back, and its azimuthal
decorrelation maps the target unintegrated glue, third there is a smooth
matching, at the leading--log($p_\perp^2$)--level to the NLO of collinear 
factorization. In Fig. 1 we compare a calculation using a realistic
unintegrated glue with recent data from the STAR
Collaboration\cite{STAR}. Notice that here no initial state smearing
was included, and improvement can be expected, see
e.g. Ref. (\refcite{Antoni}), although especially at lower $p_\perp$ such
smearing  will depend on a (model dependent) extrapolation of the pQCD cross section.
A detailed discussion will be found elsewhere (\refcite{NSSpions}).

\begin{figure}[pb]
\centerline{\psfig{file=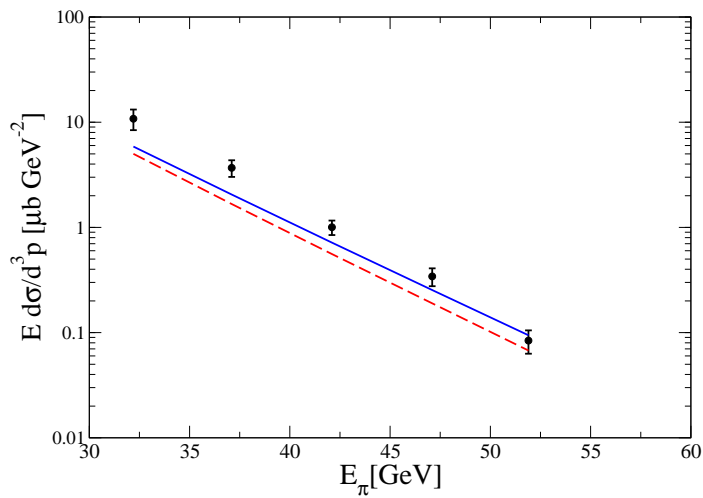,width=8.0cm,angle=270}}
\vspace*{8pt}
\caption{Invariant cross section for inclusive $\pi^0$ production
at $\sqrt{s} = 200$ GeV in pp collisions, at mean (pseudo-)rapidity $\eta = 3.8$
of pions. Data from
STAR collaboration$^6$. 
%\protect{\cite{STAR}}.  
The solid and dashed lines are for two
different sets of fragmentation functions$^{14}$ 
%\protect{\cite{Fragmentation}}, 
KKP and 
Kretzer, respectively.}
\end{figure}

\section{Nonlinear $k_\perp$--factorization}

The crucial assumption behind the linear $k_\perp$ factorization 
discussed above was that we 
could restrict ourselves to the single $t$--channel gluon exchange,
and could neglect the constraints of $s$--channel partial wave
unitarity.
Such assumption is warranted, in a limited energy range, by the finding
of only a small fraction of diffractive DIS--events  at HERA, in the 
color dipole--proton scattering at HERA energies
$\sigma_{el}/\sigma_{tot} \ll 0.5$. Still, unitarity constraints will
 be looming at LHC energies, even in $pp$ collisions, and the 
fate of $k_\perp$--factorization in a regime of strong absorption
is an important issue. It is then particularly fortunate that heavy
nuclei provide a possible testing ground for unitarity effects.
The latter will be enhanced by the size/opacity of a nucleus
and are controlled by a new large parameter, 
the saturation scale $Q_A^2 \propto A^{1/3}$.  
A pertinent quantity in the emerging formalism is the unintegrated
glue of a nucleus, $\phi(\bf{b},\bf{\kappa})$ (here $\bf{b}$ is an impact
parameter, and $\kappa$ the transverse momentum of the gluon) which is calculable 
as an expansion over multiple convolutions of the free--nucleon glue
$ f(\bf{\kappa})$. Important features of the unintegrated glue for heavy
nuclei constructed from a realistic free--nucleon input are
\cite{NSSdijet,NonlinearDijets}: 
a) a saturation
scale $ Q_A^2 \sim 0.8 \div 1 GeV$, b) a large--$\bf{\kappa}$
Cronin--type antishadowing enhancement, c) it furnishes a \emph{linear}
$k_\perp$
factorization of inclusive deep inelastic scattering, forward single-jets in
DIS,  and diffractive dijet amplitudes, d) as shown in \refcite{RealVirtual}, the
first steps of its small-$x$ evolution are governed by a certain nonlinear
version of BFKL, the Balitsky--Kovchegov equation.
While in DIS on the free nucleon target  azimuthal decorrelations
of the forward dijets map out the proton's unintegrated glue
\cite{HERADijets}, 
we found in (\refcite{NonlinearDijets}) that linear $k_\perp$--factorization
for the dijet spectrum breaks completely in the saturation regime. As a result an extension
of $k_\perp$ factorization to a strongly absorptive nuclear 
environment has been worked out, and dubbed nonlinear $k_\perp$
factorization\cite{NonlinearDijets,Nonlinear,SingleJet,RealVirtual}.
Space limitations only allow a rough sketch of our most relevant results
for the induced radiation process of interest: for rapidities $y > \log(1/x_A) \, , \, x_A
\sim 1/ R_A m_N$ ($R_A$ is the nuclear radius, $m_N$ the proton mass),
the transition $q \to qg$ proceeds coherently over the whole nucleus.
Multiple scatterings enhanced by the nuclear size/opacity are evaluated
in a Glauber--Gribov theory with coupled color channels. Nonlinear 
$k_\perp$ factorization quadratures for all the
relevant partonic subprocesses, $q \to qg, g \to q\bar{q}, g\to gg$ have
been obtained in Refs. (\refcite{Nonlinear}). 
Here we also mention
results obtained in the Color Glass Condensate approach,
which report a similar breaking of linear $k_\perp$--factorization (for
reviews and references, see e.g. Refs.(\refcite{CGC})).
Many of the nontrivial features derive from its color coupled channel
aspect. Indeed for the breakup $q \to qg$ two reaction classes are
of relevance, the first being excitation from lower to higher color multiplet: 
$3 \otimes 8 \supset  6 + 15$. Here the cross section assumes a form
\begin{equation}
d\sigma(q \to qg(6 + 15)) \propto  \int 
\Phi_{ISI,q} \otimes d\sigma^* \otimes \Phi_{FSI,g} \otimes \Phi_{FSI,q}
\, ,
\label{Eq:Higher_Mult}
\end{equation} 
which encodes at the same time probablistic $p_\perp$--broadening
through initial and final state scattering (the relevant $
\Phi_{FSI,ISI}$ being calculable in terms of the free nucleon
unintegrated glue), as well as \emph{coherent distortions} of the 
$q \to qg$ transition, which enter the effective cross section
$d\sigma^*$.
The second reaction universality class is a rotation within the initial
state multiplet, $  3 \otimes 8 \supset  3 $, here we have:

\begin{equation}
d\sigma (q \to qg(3) ) \propto \phi({\bf{b}}, {\bf{\kappa}}) |\psi_{dist} (z, {\bf{p}} -
{\bf{\kappa}}) - \psi(z, {\bf{p}}- z\bf{\kappa})|^2 \, .
\label{Eq:Triplet}
\end{equation}

There is a prefactor $ \phi({\bf{b}}, {\bf{\kappa}})$, but the
dependence on $\phi$ is highly nonlinear, because of 
the coherently distorted light--cone wavefunction of the 
$q \to qg$ transition. Furthermore here the collinear pole of
the $q \to qg$ splitting is manifest in the second undistorted
wavefunction, and we may interpret it in terms of a nuclear 
modification of the quark fragmentation.
Although it has not been a part of the original presentation, we 
mention, that equations (\ref{Eq:Higher_Mult},\ref{Eq:Triplet}) can
be given a simple Reggeon field theory interpretation in the spirit 
of AGK rules, and can be extended to more refined observables, such as
topological cross sections \cite{Cutting}.  These results
will be important for the calculation of the quenching 
of forward jets as observed in $dA$ collisions at RHIC. 

%\begin{thebibliography}{000} %for 3 digits
%\begin{thebibliography}{00}  %for 2 digits


\begin{thebibliography}{0}    %for 1 digit
 
\bibitem{Collinear}
  J.~F.~Owens,
 %  ``Large Momentum Transfer Production Of Direct Photons, Jets, And
  %Particles,''
  Rev.\ Mod.\ Phys.\  {\bf 59}, 465 (1987);
  %%CITATION = RMPHA,59,465;%%
E. Leader and E. Predazzi, Introduction to Gauge Theories
and Modern Particle Physics, vols.1 \& 2, Cambridge University Press,
Cambridge, 1996; G. Sterman, An Introduction to Quantum Field
Theory, Cambridge University Press, Cambridge, 1993; 
R.~K.~Ellis, W.~J.~Stirling and B.~R.~Webber,
``QCD and collider physics,''
Cambridge Monogr.\ Part.\ Phys.\ Nucl.\ Phys.\ Cosmol.\  {\bf 8}, 1 (1996).

\bibitem{Ktfac}
B.~Andersson {\it et al.}  [Small x Collaboration],
%``Small x phenomenology: Summary and status,''
Eur.\ Phys.\ J.\ C {\bf 25}, 77 (2002);
J.~R.~Andersen {\it et al.}  [Small x Collaboration],
%``Small x phenomenology: Summary and status 2002,''
Eur.\ Phys.\ J.\ C {\bf 35}, 67 (2004); arXiv:hep-ph/0604189;
J.~Collins and H.~Jung,
  %``Need for fully unintegrated parton densities,''
  arXiv:hep-ph/0508280.

\bibitem{Antoni}
A.~Szczurek,
%``Unintegrated parton distributions and particle production in hadronic
%collisions,''
Acta Phys.\ Polon.\ B {\bf 35}, 161 (2004);  
M.~Czech and A.~Szczurek,
  %``Unintegrated CCFM parton distributions and pion production in proton proton
  %collisions at high energies,''
Phys.\ Rev.\ C {\bf 72}, 015202 (2005); J. Phys. G {\bf 32}, 1253 (2006).
%%CITATION = HEP-PH 0311175;%%
%\cite{Nikolaev:2004cu}
\bibitem{SingleJet}
  N.~N.~Nikolaev and W.~Sch\"afer,
  %``Breaking of k(T)-factorization for single jet production off nuclei,''
  Phys.\ Rev.\ D {\bf 71}, 014023 (2005).
%  [arXiv:hep-ph/0411365].
  %%CITATION = HEP-PH 0411365;%%

%\cite{Nikolaev:2004cu}
\bibitem{RealVirtual}
  N.~N.~Nikolaev and W.~Sch\"afer,
  %``Breaking of k(T)-factorization for single jet production off nuclei,''
  Phys.\ Rev.\ D {\bf 74}, 014023 (2006).
%  [arXiv:hep-ph/0411365].
  %%CITATION = HEP-PH 0411365;%%

%\cite{Adams:2003fx}
\bibitem{STAR}
  J.~Adams {\it et al.}  [STAR Collaboration],
  %``Cross sections and transverse single-spin asymmetries in forward  neutral
  %pion production from proton collisions at s**(1/2) = 200-GeV,''
  Phys.\ Rev.\ Lett.\  {\bf 92}, 171801 (2004).
 
\bibitem{NSSpions}
N.~N.~Nikolaev, W.~Sch\"afer and  A.~Szczurek,
in preparation.

\bibitem{NSSdijet}
N.~N.~Nikolaev, W.~Sch\"afer and G.~Schwiete,
%``Coherent production of hard dijets on nuclei in QCD,''
Phys.\ Rev.\ D {\bf 63}, 014020 (2001).

\bibitem{NonlinearDijets}
N.~N.~Nikolaev, W.~Sch\"afer, B.~G.~Zakharov and V.~R.~Zoller,
%``Nonlinear k(T) factorization for forward dijets in DIS off nuclei in  the
%saturation regime,''
J.\ Exp.\ Theor.\ Phys.\  {\bf 97}, 441 (2003);
[Zh.\ Eksp.\ Teor.\ Fiz.\  {\bf 124}, 491 (2003)]
%[arXiv:hep-ph/0303024].
%%CITATION = HEP-PH 0303024;%%{arXiv: hep-ph/0303024}

\bibitem{HERADijets}
  A.~Szczurek, N.~N.~Nikolaev, W.~Sch\"afer and J.~Speth,
%``Mapping the proton unintegrated gluon distribution in dijets  correlations
%in real and virtual photoproduction at HERA,''
  Phys.\ Lett.\ B {\bf 500}, 254 (2001).
  %%CITATION = HEP-PH 0011281;%%

%\cite{Nikolaev:2005qs}
\bibitem{Nonlinear}
  N.~N.~Nikolaev, W.~Sch\"afer and B.~G.~Zakharov,
  %``Nonuniversality aspects of nonlinear k(T)-factorization for hard  dijets,''
  Phys.\ Rev.\ Lett.\  {\bf 95}, 221803 (2005); Phys.\ Rev.\ D {\bf 72},
  114018 (2005);  N.~N.~Nikolaev, W.~Sch\"afer, B.~G.~Zakharov and V.~R.~Zoller,
  %``Nonlinear k(T)-factorization for quark-gluon dijet production off
  %nuclei,' 
 Phys.\ Rev.\ D {\bf 72}, 034033 (2005).
%%CITATION = HEP-PH 0502018;%%

%\cite{Jalilian-Marian:2005jf}
\bibitem{CGC}
  J.~Jalilian-Marian and Y.~V.~Kovchegov,
  %``Saturation physics and deuteron gold collisions at RHIC,''
  Prog.\ Part.\ Nucl.\ Phys.\  {\bf 56}, 104 (2006);
  H.~Fujii, F.~Gelis and R.~Venugopalan,
  %``Quark production in high energy proton nucleus collisions,''
  Eur.\ Phys.\ J.\ C {\bf 43}, 139 (2005).
  %%CITATION = HEP-PH 0502204;%%
  %%CITATION = HEP-PH 0505052;%%

%\cite{Nikolaev:2006mx}
\bibitem{Cutting}
  N.~N.~Nikolaev and W.~Sch\"afer,
  %``Unitarity cutting rules for the nucleus excitation and topological cross
  %sections in hard production off nuclei from nonlinear k(T)-factorization,''
  arXiv:hep-ph/0607307.
  %%CITATION = HEP-PH 0607307;%%

 
%\cite{Kniehl:2000fe}
\bibitem{Fragmentation}
  B.~A.~Kniehl, G.~Kramer and B.~Potter,
  %``Fragmentation functions for pions, kaons, and protons at  next-to-leading
  %order,''
  Nucl.\ Phys.\ B {\bf 582}, 514 (2000);  S.~Kretzer,
%   ``Fragmentation functions from flavour-inclusive and flavour-tagged e+ e-
%annihilations,''
  Phys.\ Rev.\ D {\bf 62}, 054001 (2000).
  %%CITATION = HEP-PH 0010289;%%

\end{thebibliography}
\end{document}